\documentclass[]{aastex631}

\shorttitle{A Standard FRB VOEvent Service}

\graphicspath{{./}{figures/}}

\begin{document}

\title{\texttt{frb-voe}: A Real-time Virtual Observatory Event Alert Service for Fast Radio Bursts}

\correspondingauthor{Thomas C. Abbott, Andrew V. Zwaniga}
\email{thomas.abbott@mail.mcgill.ca, azwaniga@torontomu.ca}

\author[0000-0001-5002-0868]{Thomas C. Abbott}
\affiliation{Department of Physics, McGill University, 3600 rue University, Montr\'eal, QC H3A 2T8, Canada}
\affiliation{Trottier Space Institute at McGill University, 3550 rue University, Montr\'eal, QC H3A 2A7, Canada}

\author[0000-0001-8278-1936]{Andrew V. Zwaniga}
\affiliation{Department of Physics, Toronto Metropolitan University, 350 Victoria St, Toronto ON M5B 2K3, Canada}

\author[0000-0002-1800-8233]{Charanjot Brar}
\affiliation{Department of Physics, McGill University, 3600 rue University, Montr\'eal, QC H3A 2T8, Canada}
\affiliation{Trottier Space Institute at McGill University, 3550 rue University, Montr\'eal, QC H3A 2A7, Canada}

\author[0000-0001-9345-0307]{Victoria M. Kaspi}
\affiliation{Department of Physics, McGill University, 3600 rue University, Montr\'eal, QC H3A 2T8, Canada}
\affiliation{Trottier Space Institute at McGill University, 3550 rue University, Montr\'eal, QC H3A 2A7, Canada}

\author[0000-0002-9822-8008]{Emily Petroff}
\affiliation{Department of Physics, McGill University, 3600 rue University, Montr\'eal, QC H3A 2T8, Canada}
\affiliation{Trottier Space Institute at McGill University, 3550 rue University, Montr\'eal, QC H3A 2A7, Canada}
\affiliation{Perimeter Institute for Theoretical Physics, 31 Caroline Street N, Waterloo, ON N25 2YL, Canada}

\author[0000-0002-3615-3514]{Mohit Bhardwaj}
\affiliation{McWilliams Center for Cosmology, Department of Physics, Carnegie Mellon University, Pittsburgh, PA 15213, USA}

\author[0000-0001-8537-9299]{P.J. Boyle}
\affiliation{Department of Physics, McGill University, 3600 rue University, Montr\'eal, QC H3A 2T8, Canada}

\author[0000-0001-6422-8125]{Amanda M. Cook}
\affiliation{David A. Dunlap Institute Department of Astronomy \& Astrophysics, University of Toronto, 50 St. George Street, Toronto, Ontario, Canada M5S 3H4}
\affiliation{Dunlap Institute for Astronomy \& Astrophysics, University of Toronto, 50 St.~George Street, Toronto, ON M5S 3H4, Canada}

\author[0000-0003-3457-4670]{Ronniy C. Joseph}
\affiliation{Department of Physics, McGill University, 3600 rue University, Montr\'eal, QC H3A 2T8, Canada}
\affiliation{Trottier Space Institute at McGill University, 3550 rue University, Montr\'eal, QC H3A 2A7, Canada}

\author[0000-0002-4279-6946]{Kiyoshi W. Masui}
\affiliation{MIT Kavli Institute for Astrophysics and Space Research, Massachusetts Institute of Technology, 77 Massachusetts Ave, Cambridge, MA 02139, USA}
\affiliation{Department of Physics, Massachusetts Institute of Technology, 77 Massachusetts Ave, Cambridge, MA 02139, USA}

\author[0000-0002-8897-1973]{Ayush Pandhi}
\affiliation{David A. Dunlap Institute Department of Astronomy \& Astrophysics, University of Toronto, 50 St. George Street, Toronto, Ontario, Canada M5S 3H4}
\affiliation{Dunlap Institute for Astronomy \& Astrophysics, University of Toronto, 50 St.~George Street, Toronto, ON M5S 3H4, Canada}

\author[0000-0002-4795-697X]{Ziggy Pleunis}
\affiliation{Anton Pannekoek Institute for Astronomy, University of Amsterdam, Science Park 904, 1098 XH, Amsterdam, The Netherlands.}
\affiliation{ASTRON, Netherlands Institute for Radio Astronomy, Oude Hoogeveensedijk 4, 7991 PD Dwingeloo, The Netherlands.}

\author[0000-0002-7374-7119]{Paul Scholz}
\affiliation{Department of Physics and Astronomy, York University, 4700 Keele Street, Toronto, Ontario, ON MJ3 1P3, Canada}
\affiliation{Dunlap Institute for Astronomy \& Astrophysics, University of Toronto, 50 St.~George Street, Toronto, ON M5S 3H4, Canada}
\author[0000-0002-6823-2073]{Kaitlyn Shin}
\affiliation{MIT Kavli Institute for Astrophysics and Space Research, Massachusetts Institute of Technology, 77 Massachusetts Ave, Cambridge, MA 02139, USA}
\affiliation{Department of Physics, Massachusetts Institute of Technology, 77 Massachusetts Ave, Cambridge, MA 02139, USA}

\author[0000-0003-2548-2926]{Shriharsh Tendulkar}
\affiliation{Department of Astronomy and Astrophysics, Tata Institute of Fundamental Research, Mumbai, 400005, India}
\affiliation{National Centre for Radio Astrophysics, Post Bag 3, Ganeshkhind, Pune, 411007, India}

\begin{abstract}

We present \texttt{frb-voe}, a publicly available software package that enables radio observatories to broadcast fast radio burst (FRB) alerts to subscribers through low-latency virtual observatory events (VOEvents). We describe a use-case of \texttt{frb-voe} by the Canadian Hydrogen Intensity Mapping Experiment Fast Radio Burst (CHIME/FRB) Collaboration, which has broadcast thousands of FRB alerts to subscribers worldwide. Using this service, observers have daily opportunities to conduct rapid multi-wavelength follow-up observations of new FRB sources. Alerts are distributed as machine-readable reports and as emails containing FRB metadata, and are available to the public within approximately 13 seconds of detection. A sortable database and a downloadable JSON file containing FRB metadata from all broadcast alerts can be found on CHIME/FRB's public webpage. The \texttt{frb-voe} service also provides users with the ability to retrieve FRB names from the Transient Name Server (TNS) through the \texttt{frb-voe} client user interface (CLI). The \texttt{frb-voe} service can act as a foundation on which any observatory that detects FRBs can build its own VOEvent broadcasting service to contribute to the coordinated multi-wavelength follow-up of astrophysical transients. 

\end{abstract}

\section{Introduction} 
\label{sec:intro}

Fast radio bursts (FRBs) are brief ($\mu$s -- ms), energetic astrophysical events originating from extragalactic sources with unknown physical origins (for a review, see \citealt{Petroff+2022}). Since their discovery in 2007 \citep{Lorimer_2007}, approximately 1000 FRB sources have been detected by dozens of observatories across the globe\footnote{See the TNS database: \url{https://www.wis-tns.org/}}. A large fraction of those detections have been made by the Canadian Hydrogen Intensity Mapping Experiment (CHIME), located in Penticton, British Columbia, Canada.
CHIME is a transit radio telescope consisting of four semi-cylindrical parabolic reflectors with a $400 - 800$ MHz receiving band. The telescope houses several data processing ``backends" including a backend dedicated to searching for FRBs, known as the CHIME/FRB backend. In the first CHIME/FRB Catalog, the CHIME/FRB Collaboration published over 500 detections of new FRB sources \citep{CHIMEFRB_Catalog1_2021}. With a growing number of active FRB-observatories and FRB detections, there is a need for a fast, standardized, and concise way to communicate between observatories, as well as infrastructure that efficiently enables this process. In addition, many FRB progenitor models predict a multi-wavelength counterpart signal directly following an FRB (e.g., \citealt{Shand+2016}, \citealt{Wang+2016}, \citealt{Margalit+2019}) suggesting that low-latency follow-up could constrain and even rule out or provide strong evidence in favour of some of these models. Prior to public alerts, \cite{CHIMEFRB_SGR_2020} published the detection of an FRB-like signal from the Galactic magnetar SGR 1935+2154, which, together with the simultaneous X-ray activity (\citealt{Mereghetti+2020}) makes this a landmark result and lends credence to the pursuit of multi-wavelength follow-up of other FRBs. A search for multi-wavelength, and/or multi-messenger, counterparts of a population of FRBs calls for a system in which CHIME/FRB provides alerts to observers having a variety of follow-up resources. Hence, we introduce a real-time FRB public alert system hosted at CHIME, known as the CHIME/FRB Virtual Observatory Event (VOEvent) Service, as well as a telescope agnostic FRB public alert system, known as \texttt{frb-voe}. The \texttt{frb-voe} codebase is available for download and open for contributions by motivated community members\footnote{\url{https://github.com/CHIMEFRB/voe}} and the CHIME/FRB Virtual Observatory Event Service is actively contributing to low-latency triggered multi-wavelength FRB follow-up efforts.

The \texttt{frb-voe} service follows the FRB VOEvent Standard, which was first introduced by \cite{Petroff+2017} and provides a standardized format for communicating FRB observations to automated observatories around the world over the internet. A VOEvent is a standard information packet for communicating a transient celestial event, especially for rapid follow-up \citep{Seaman+2011}. In general, VOEvents standardize \textit{how} an observatory should report an observation, leaving flexibility for the author to decide \textit{what} scientific metadata that report should contain. As an example, the metadata included in the CHIME/FRB VOEvent Service's alerts are presented in Table \ref{tab:metadata}. Each FRB VOEvent has a unique International Virtual Observatory Resource Name (IVORN), and an alert type that is one of \textit{detection}, \textit{subsequent}, \textit{retraction}, or \textit{update}. The other parameters reported in the VOEvent include basic metadata products from the real-time FRB detection pipeline, and the characteristics of the FRB detection pipeline. For the CHIME/FRB VOEvent Service, full details of how these metadata are produced from the pipeline are available in \cite{CHIMEFRB_Overview_2018}.

\begin{table}
    \centering
    \begin{tabular}{|l|l|l|l}
        \textbf{Item} & \textbf{Description} & \textbf{Unit} & \textbf{Example for CHIME/FRB} \\
        \texttt{ivorn} & VOEvent  International Virtual Observatory Resource Name & - & ivo://ca.chimenet.frb/...\\
        Alert Type & One of: \textit{detection}, \textit{subsequent}, \textit{retraction}, or \textit{update} & - & \textit{detection} \\
        \texttt{sampling\_time}  & FRB search temporal resolution & ms & 0.983 \\
        \texttt{bandwidth}  & Host observatory frequency range & MHz & 400 \\ 
        \texttt{centre\_frequency}  & Host observatory frequency range midpoint & MHz & 600 \\ 
        \texttt{npol}  & Number of polarizations summed in FRB search& - & 2 \\ 
        \texttt{bits\_per\_sample}  & FRB search time sample bit count & - & 8 \\ 
        \texttt{tsys}  & Host observatory receiver noise temperature & K & 50 \\ 
        \texttt{backend}  & Processing backend & - & CHIME/FRB backend \\ 
        \texttt{event\_no}  & VOEvent database primary key & - & 1234567890 \\ 
        \texttt{known\_source\_name}  & FRB source name (a TNS name, if available) & - & FRB20180916B \\ 
        \texttt{event\_type}  & Status of FRB source identification & - & \{known, unknown\} \\ 
        \texttt{pipeline\_name}  & real-time FRB pipeline & - & CHIME/FRB Pipeline \\ 
        \texttt{dm}  & Radio signal dispersion measure & pc/cc & 547.1\footnote{\label{foot: table-uncertainty}The corresponding uncertainty on these values is also in the VOEvent under the same name but appended with \texttt{\_error}.} \\ 
        \texttt{timestamp\_utc}  & UTC topocentric arrival time & datetime & 2022-03-28 21:35:18.932183\textsuperscript{\ref{foot: table-uncertainty}} \\ 
        \texttt{snr}  & Detection signal-to-noise ratio & - & 20.0 \\ 
        \texttt{pos\_semiminor\_deg\_95}  & Localization ellipse semi-minor axis & deg. & 0.48 \\ 
        \texttt{pos\_semimajor\_deg\_95}  & Localization ellipse semi-major axis & deg. & 0.71 \\ 
        \texttt{dm\_gal\_ne\_2001\_max}  & NE 2001 Galactic DM Model estimate & pc/cc & 30.5 \\ 
        \texttt{dm\_gal\_ymw\_16\_max}  & YMW 2016 Galactic DM Model estimate & pc/cc & 22.4 \\ 
        \texttt{timestamp\_utc\_inf\_freq}  & UTC topocentric arrival time at infinite frequency & datetime & 2022-06-14 03:20:09.017604 \\ 
        Coordinate System  & Astrophysical coordinate system & - & UTC-FK5-TOPO \\ 
        Localization  & Localization ellipse central coordinates\footnote{The localization provided in CHIME/FRB VOEvents is derived from the central lobe of the real-time system's ``header" localization (see Figure 6 of \citep{CHIMEFRB_Catalog1_2021} for examples.} & deg. & (191.84, 1.91) \\ 
        Links  & Public webpage & - & \url{https://www.chime-frb.ca} \\ 
        Importance  & Machine learning score from 0 (false positive) to 1 & - & 0.98 \\ 
        Probability & Probability of known source association & - & 0.23${}^{*}$ \\ 
        Citations  & IVORNs of one or more cited VOEvents & - & ivo://ca.chimenet.frb/... \\ 
    \end{tabular}
    \caption{Metadata that are currently published in CHIME/FRB VOEvents, restricted to items most relevant to follow-up. Names styled in mono-space font indicate metadata that appear within the VOEvent XML document as a \texttt{Param}. Names styled in the common font are not a \texttt{Param} but rather are subsections named as such within the XML. We note that the VOEvent type can be found within the IVORN string, however, we recommend also including the VOEvent type as a its own parameter in the VOEvent, for subscriber convenience.} \label{tab:metadata}
\end{table}

In addition to reporting metadata according to the FRB VOEvent Standard, an important step in harmonizing the reporting of FRB detections taken by the community is the adoption of a naming convention for FRBs officially recognized by the International Astronomical Union (IAU). The Transient Name Server (TNS) now supplies and maintains the official FRB naming scheme and users can submit their FRB discoveries using a graphical user interface on the TNS webpage, or programmatically\footnote{See section on APIs at \url{https://www.wis-tns.org/content/tns-getting-started}} using various scripting languages like Python, to receive a name for the FRB. Through the TNS webpage, or in some cases programmatically, users can also query existing FRBs and cross-reference with other astrophysical transients that are curated by the server. The TNS FRB naming scheme consists of the discovery date in YYYYMMDD format, and proceeds alphabetically according to the submission time to the TNS. For example, two FRBs discovered on January 2, 2023 would be named FRB 20230102A and FRB 20230102B such that the submission time of A precedes B. Another important note is that repeating FRB sources are typically referred to by their \textit{source name}, i.e. the TNS name corresponding to their first detection, and all subsequent detections of the source are given their own TNS name but are used only to uniquely identify specific bursts from repeaters. Finally, the source name is not intended to imply that the source turned on or became active on that date; rather, it indicates when it was first detected. This naming scheme was used in the publication of the first catalog of FRBs from CHIME/FRB (\citealt{CHIMEFRB_Catalog1_2021}) and allowed researchers around the world communicate coherently about individual FRB sources.

This paper presents a standard telescope-agnostic FRB VOEvent alert system, which is publicly available in an open-source software package called \texttt{frb-voe} and provides a framework for all FRB observatories to share coordinated real-time observations. In Section \ref{sec:code}, we describe the core components of \texttt{frb-voe}, including VOEvent authoring, an integrated Comet VOEvent broker\footnote{\url{https://comet.transientskp.org/en/stable/}} (\citealt{Swinbank2014}), and routines for programmatically managing an observatory's FRB discoveries on the TNS. In Section \ref{sec: chime-voevents}, we discuss science use-cases of \texttt{frb-voe}, focusing on CHIME/FRB as an example and in Section \ref{sec:discussion}, we offer motivation to other radio observatories to adopt an FRB VOEvent system. In Section \ref{sec:conclude}, we offer concluding remarks and look at the future of FRBs and VOEvents.

\section{frb-voe \label{sec:code}}

\subsection{Software}
\begin{figure}
    \centering
    \includegraphics[scale=0.55]{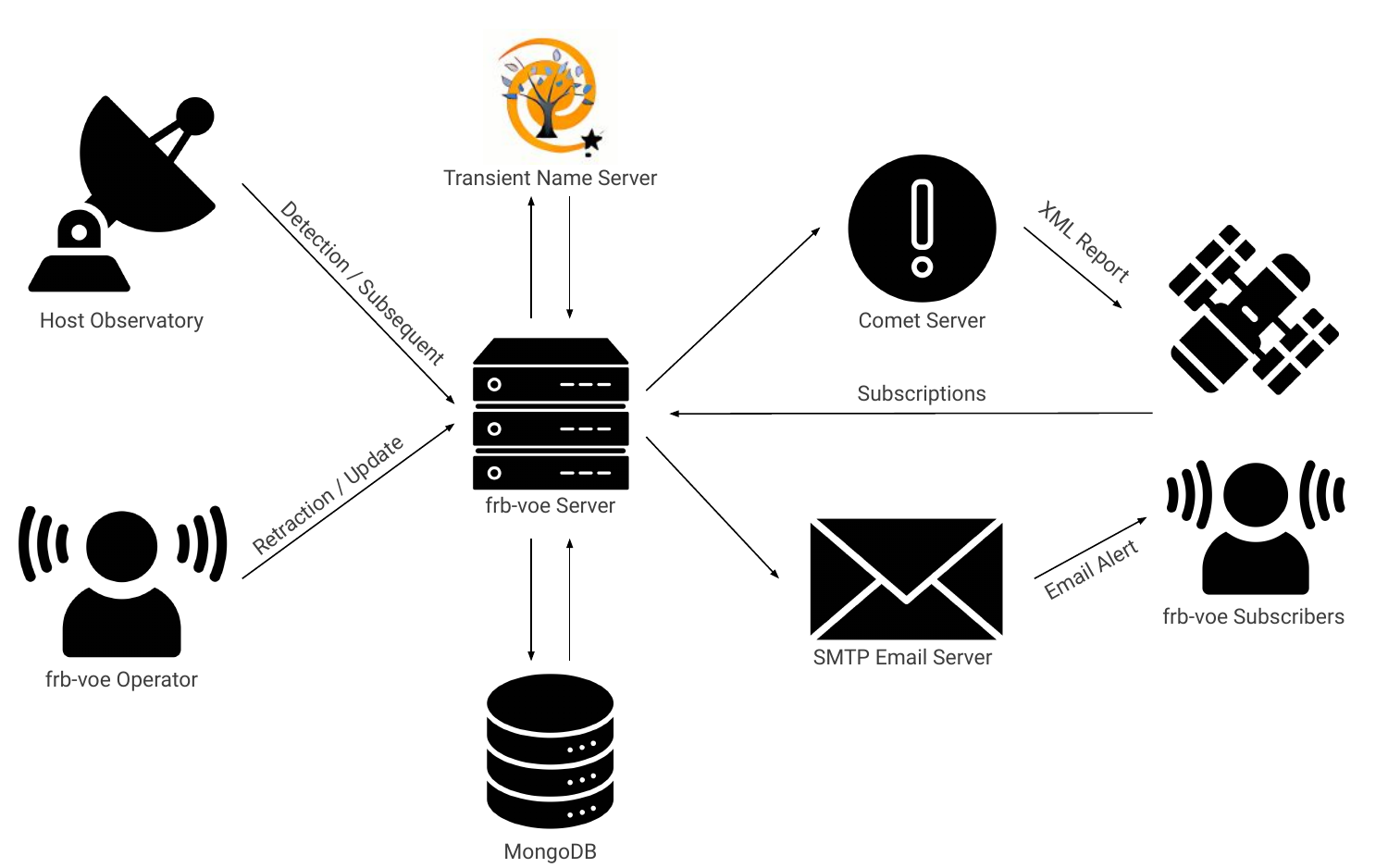}
    \caption{Workflow for the FRB alert service. Observatories that detect FRBs in real-time or post real-time (e.g. archival searches) can make equal use of this workflow. From left to right: (upper) a burst is detected at the host observatory, either a new FRB or a burst from a known repeating FRB or (lower) the \texttt{frb-voe} service's operator initiates a retraction or update VOEvent. The metadata are sent to the \texttt{frb-voe} server via requests and then saved to the MongoDB database. Optionally, the \texttt{frb-voe} server can request a TNS name. Finally, after validating the data, the server broadcasts email and machine-readable XML alerts to subscribers and automated follow-up observatories. Icons from \href{https://thenounproject.com/}{The Noun Project} (see acknowledgments for complete list of creators). \label{fig:detection}}
\end{figure}

This Section provides details of each component of the \texttt{frb-voe} software and their function. A depiction of the workflow is presented in Figure \ref{fig:detection}. The software is available both on Github and on Zenodo\footnote{\url{https://zenodo.org/records/14008387}}, where a citable DOI can be found. The \texttt{frb-voe} software package is a REpresentational State Transfer (REST) Application Programming Interface (API) based on the \texttt{sanic}\footnote{\url{https://sanic.dev/en/}} library, which is itself a framework for building performant software servers. We note that while many software services consist of both a backend and a frontend, the present release of \texttt{frb-voe} is only a backend, which can be interacted with through HTTP requests. To trigger a VOEvent, the host observatory or the \texttt{frb-voe} operator sends an HTTP request to the \texttt{frb-voe} server. The data within the request is automatically validated using custom data models built using the \texttt{Pydantic} library (\citealt{Colvin_Pydantic_2024}). FRB VOEvents can be created as one of six alert types, as prescribed in \cite{Petroff+2017}: \textit{detection}, for new FRBs; \textit{subsequent}, for bursts from known FRBs; \textit{update}, for new measurements that supersede initial values reported in a previously broadcast VOEvent; \textit{retraction}, to mark a previously broadcast VOEvent that has been found to be a non-FRB (e.g. a false positive due to radio frequency interference (RFI)), \textit{search}, to report the start of a new blind search, and \textit{targeted} to report the start of a targetted follow-up observation. In the software presented in this paper, we offer support for the alert types reporting new information, namely, \textit{detection}, \textit{subsequent}, \textit{update}, and \textit{retraction}, however, new alert types can be added in future work. Typically, \textit{detection} and \textit{subsequent} events are triggered by the host observatory's FRB search pipeline, where as \textit{update} and \textit{retraction} events are triggered by the \texttt{frb-voe} operator; however, some exceptions include archival searches, or an FRB candidate being promoted to an FRB, in which case the \texttt{frb-voe} operator may decide to trigger a \textit{detection} VOEvent. \texttt{frb-voe} is designed with data validation in mind: a class for each alert type ensures that an FRB VOEvent is always created with a set of consistent core components. However, some flexibility remains; e.g. users can report items in addition to the core components in the VOEvent, such as observatory-specific metadata and other measured parameters. After the data is automatically validated and converted to the FRB VOEvent Standard, the event is saved to a database. MongoDB\footnote{\url{https://github.com/mongodb/mongo}} is used to instantiate database tables that store information about broadcast VOEvents and VOEvent subscribers. Optionally, the VOEvent can be sent to the TNS to acquire a TNS name. Finally, a connection to a local Comet VOEvent broker (\citealt{Swinbank2014}) allows the VOEvent to be broadcast to a subscriber list loaded from the database. For subscribers that opt in to email alerts, a request is also sent to a simple mail transfer protocal (SMTP) email server to be broadcast. Altogether, the system can accomplish tasks such as creating FRB VOEvents, broadcasting the VOEvent to subscribers, and adding an entry to the TNS to obtain the standard name for the FRB (see Figure \ref{fig:detection}). 

\subsection{Installation \label{subsec: installation}}
The \texttt{frb-voe} service is designed to be installed and operated in a \texttt{Docker} container. \texttt{Docker} is a software platform used to build containerized applications, enabling the \texttt{frb-voe} service to be run on any machine. The user should select a dedicated machine on which to install the code and run the backends, then, proceed with the installation instructions found on GitHub\footnote{\url{https://github.com/CHIMEFRB/voe}}. Upon building the \texttt{frb-voe} container, each component of the \texttt{frb-voe} system is automatically initiated and ready for use. If the operator would like to broadcast email alerts, a connection to an SMTP server will be necessary. 

\subsection{Using the Code \label{subsec:code-usage}}

\begin{figure}[h!]
    \centering
    \includegraphics[scale=0.5]{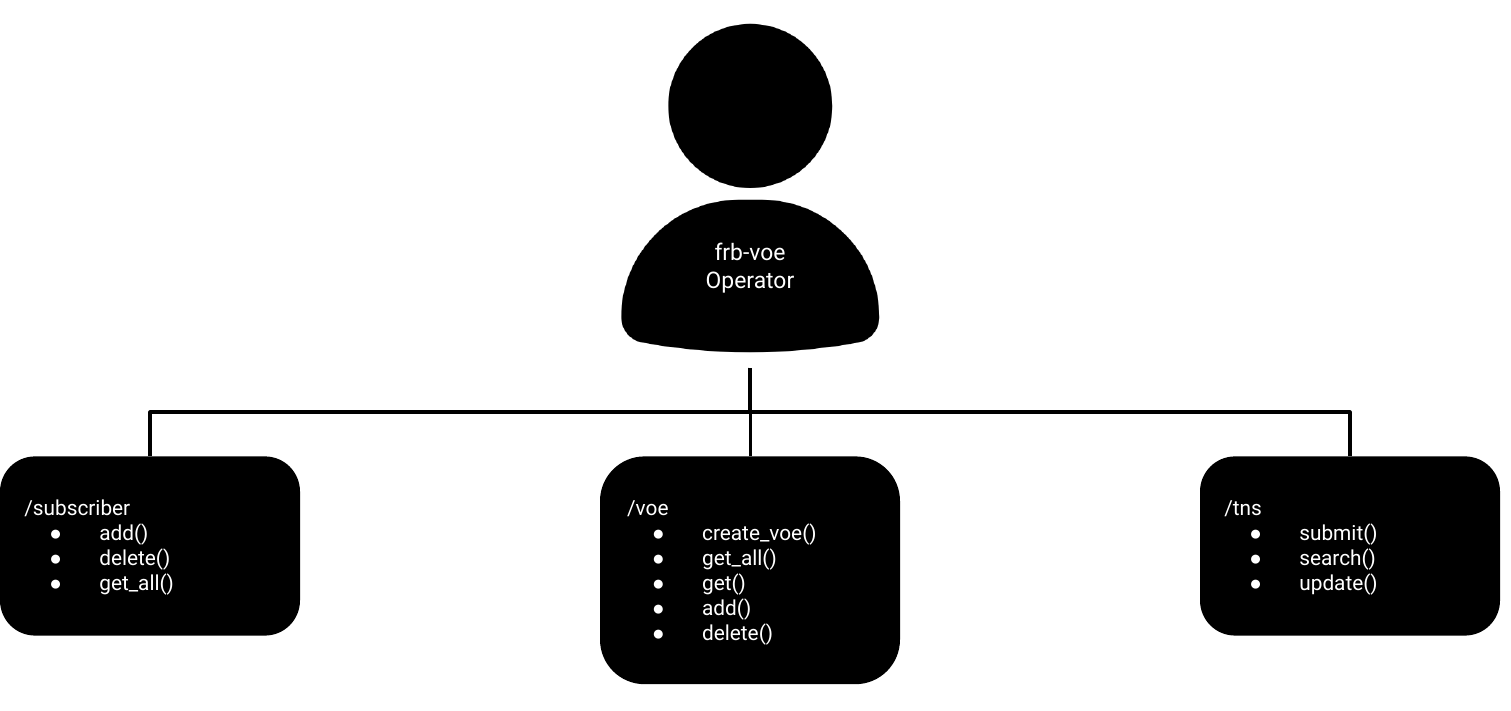}
    \caption{The three interactable backends (tns, voe, and subscriber) available to the operator of \texttt{frb-voe}. The backends can be accessed using a built-in CLI. As an example, publishing a VOEvent \ref{fig:detection} is done by an HTTP request with URL set to \texttt{https://localhost:port/frb-voe/voe/create\_voe}. \label{fig:backends}}
\end{figure}

Interaction with the backend is initiated through HTTP requests; via the \texttt{requests}\footnote{\url{https://pypi.org/project/requests/}} package. To broadcast a VOEvent, a request must be sent from the host observatory or \texttt{frb-voe} operator to the \texttt{frb-voe} server, an example script to do so is provided in the Github repository. For the three backends shown in Figure \ref{fig:backends}, a requests can be sent in a standard way using a command line interface (CLI), via the \texttt{click}\footnote{\url{https://click.palletsprojects.com/en/8.1.x/}} package. Each request is passed to a routine (shown as bullet points in Figure \ref{fig:backends}) which performs a task for the user, such as a calculation or transformation of the input. Input and output are handled in JavaScript Object Notation (JSON), which is processed in Python via dictionaries and the built-in \texttt{json} library. In general, requests are sent to a host machine at an address designated with a central URL, e.g., \texttt{https://foo.bar/frb-voe}. Individual routines are selected by sending request to sub-paths of this URL, e.g. \texttt{https://foo.bar/frb-voe/voe}.

As shown in Figure \ref{fig:backends}, each backend offers several routines with functionalities that enable use and maintenance of the \texttt{frb-voe} system. The Subscribers backend allows adding or deleting entries to a subscriber database through a CLI. This has both operational and scientific advantages: firstly, the host observatory gains visibility into what applications their public alerts are used for: and secondly, it is useful to initially recruit a set of temporary test subscribers before inviting the general public to subscribe. Underneath the hood, the subscriber database is loaded by the Comet broker upon initializing the broker. The TNS backend allows the user to (1) submit an FRB to the TNS, which then responds with the official TNS name for the burst; (2) search for an existing FRB object on the TNS; and (3) update the proprietary period of an existing proprietary FRB object on the TNS. Lastly, the VOE backend allows the user to publish any of the four FRB VOEvent types to the Comet broker or to a list of email addresses. Every VOEvent is saved to an underlying database that can be queried through the VOE backend CLI. 

Many of the routines in Figure \ref{fig:backends} require JSON payloads with specific key-value pairs. For example, the \texttt{/tns/submit} routine takes a payload of data with a specific format and then creates an FRB JSON report to be submitted to the TNS. Custom data classes within \texttt{frb-voe} ensure the payload into \texttt{/tns/submit} must follow the standard defined in \texttt{frb-voe} so that the data will be accepted by the external TNS API,. Similarly, the endpoints for creating FRB VOEvents (\texttt{/voe/create\_voe}, etc.) also take as input a payload with specific key-value pairs. This data validation ensures that all observatories that use \texttt{frb-voe} will publish VOEvents with consistent structure and metadata.

\section{The CHIME/FRB VOEvent Service \label{sec: chime-voevents}}

\subsection{CHIME/FRB Pipeline Overview}
The CHIME/FRB VOEvent Service, active since October 2021, is currently providing subscribers\footnote{Subscriptions are free and can be requested at \url{https://www.chime-frb.ca/voevents}.} at dozens of research institutions with triggers for facilitating multi-wavelength follow-up of newly discovered FRBs moments after the detection at CHIME. The following is a brief summary of the process that leads to triggering VOEvents;  full details are available in \cite{CHIMEFRB_Overview_2018}. The real-time FRB detection pipeline is a four level process that (1) detects dispersed radio signals and estimates the dispersion measure (DM), (2) uses a machine learning algorithm to classify the signal as \textit{RFI} or \textit{Astrophysical}, (3) localizes the signal to a region on the sky and estimates whether the signal is \textit{Extragalactic}, \textit{Galactic}, or \textit{Ambiguous} depending on a comparison to the Galactic contribution to the DM along the line-of-sight to the region's center, and (4) identifies \textit{Events} as signals that are interesting based on a combination of cuts placed on the signal-to-noise ratio (SNR) and the score of an additional algorithm provides a probability that the event originates from a known source. These decisions and calculations are stored as metadata and are passed through each successive level of the pipeline in the form of an \textit{Event} object.

When the parameters of the \textit{Event} pass certain thresholds, the final level of the pipeline triggers the VOEvent Service. Currently, the VOEvent triggering threshold is the same as that used for saving the raw voltages of the radio signal time series (at the time of writing this paper, that is a signal-to-noise threshold of 12), and this criterion generates a few alerts per day while limiting the number of retracted events. When the \textit{Event} data can be matched in real-time to a known FRB source, the alert is published as a subsequent-type VOEvent; otherwise, as a detection-type VOEvent. Note that these alerts go out in real-time, before human verification, as suggested in Figure \ref{fig:detection}. Up to a day is required for thorough review of the \textit{Event} data by CHIME/FRB scientists, after which the \textit{Event} is either promoted to \textit{Candidate} status or identified as a false positive. In the case of a false positive, a retraction-type VOEvent is published, citing the original VOEvent. 

\subsection{Performance \& Summary}

\begin{figure}
    \centering
    \includegraphics[scale=0.7]{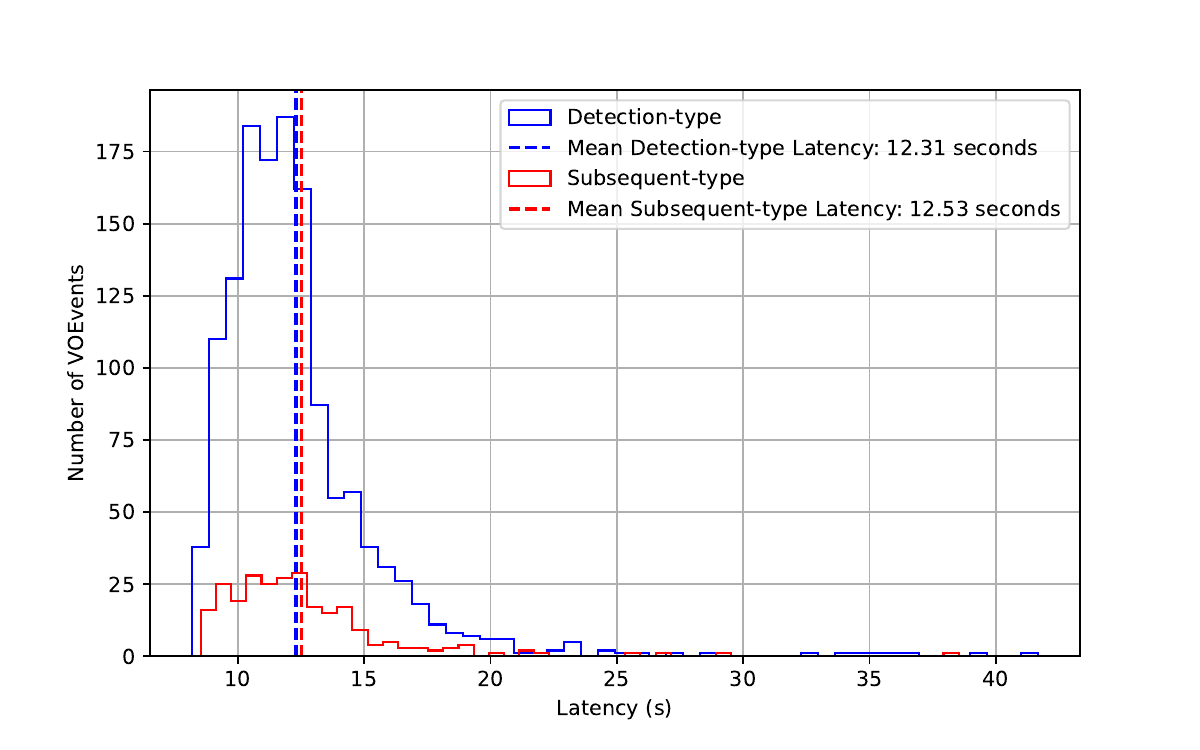}
    \caption{A Histogram of the latencies of VOEvents broadcast between the start of the service's operation (2021/10/08) and 2024/06/15. \textit{Detection} VOEvents are shown in blue, and \textit{subsequent} VOEvents are shown in red. The mean latency for each VOEvent type is shown in a dashed line and is between 12 and 13 seconds for both types. The slightly larger latency in subsequent type events may be due to an additional known source check in the real-time pipeline.  \label{fig:latency}}
\end{figure}

Between October 2021 and June 2024, the CHIME/FRB VOEvent Service has broadcast nearly 2000 FRB VOEvents of which about 1620 ($\sim$ 81\%) are detection type events, about 260 ($\sim$ 13\%) are subsequent type, and only 120 ($\sim$ 6\%) of broadcast events have been retracted. By calculating the \textit{latency}, defined by the time passed from FRB detection at 400 MHz to broadcasting the VOEvent from the Comet server, we find that for CHIME/FRB VOEvents, the value is typically less than 13 seconds. Figure \ref{fig:latency} shows the latency of CHIME/FRB VOEvents broadcast between October 2021 and June 2024. Note that the latency value comes from the postprocessing of the events in the real-time detection pipeline and is not a direct output of the \texttt{frb-voe} software. Most retraction events are broadcast within 24 hours of misdetection, the higher latency is a consequence of retraction type events being triggered by human verification. In many cases, retractions are issued for pulsars which were not correctly associated with a known source, or radio frequency interference (RFI). As of June 2024, the CHIME/FRB VOEvent Service has not broadcast update type VOEvents; however, this may change in the future if CHIME/FRB VOEvent parameters are updated through rapid refined analyses. To allow open access to the CHIME/FRB data, and therefore remove systemic barriers for all researchers around the world, a public database of all broadcast CHIME/FRB VOEvents can be found on the official CHIME website\footnote{\url{https://www.chime-frb.ca/voevents}}. The webpage also enables offline analysis of CHIME/FRB VOEvents by providing a downloadable JSON file with all FRB metadata.

\newpage    
\section{Discussion \label{sec:discussion}}

\subsection{Why use VOEvents? \label{subsec:discussion-why-voevents}}

The VOEvent format has been static since its last major release, version 2.0 (\citealt{Seaman+2011}). VOEvent XMLs are easily parsed in Python via the convenience library, \texttt{voeventparse}\footnote{\url{https://voevent-parse.readthedocs.io/en/stable/reference.html}} (\citealt{Staley2014}), making them suitable for users of all levels, from beginners to experts. As a result, VOEvents are used by dozens of observatories around the world, in particular the Gamma ray Coordinates Network (GCN/TAN)\footnote{\url{https://gcn.gsfc.nasa.gov/}}. The VOEvents distributed by GCN/TAN relate to observations of GRBs, and they follow a particular format. For instance, NASA's \emph{Swift} Telescope\footnote{\url{https://swift.gsfc.nasa.gov/}} is capable of publishing several different types of alerts relating to slewing and observations of GRBs with one of the three onboard instruments. This has allowed for the follow-up of gamma ray bursts (GRBs) by multiple observatories distributed across the world and in Earth orbit. While the flexibility in content accommodates different observation specifications, the rigidity in the structure encourages disparate observatories to communicate cooperatively and work toward the unified goal of multi-wavelength and/or multi-messenger follow-up.

\subsection{Why use \emph{\texttt{frb-voe}} ? \label{subsec:discussion-why-frb-voe}}

The \texttt{frb-voe} Service aims to provide the FRB community with a similar standard to that of GCN/TAN. The Service offers host observatories the ability to broadcast detections to a private set of approved subscribers or publicly to larger transient databases, such at the TNS. Thus far, engagement of the FRB community with CHIME/FRB VOEvents has been positive and fruitful. FRB latency has drastically decreased from day or month timescales before the implementation of FRB VOEvents to a few seconds. In addition, triggered X-ray follow-up by the \textit{Swift} BAT GUANO experiment has yielded the first Astronomer's Telegrams (ATels)\footnote{\url{https://www.astronomerstelegram.org}} based on coordinates and other metadata from a CHIME/FRB VOEvent (\citealt{2021ATel15055....1T}, \citeyear{2021ATel15114....1T}, \citeyear{2023ATel16233....1T}), and follow-up from the Deep Synoptic Array has led to the first multi-observatory FRB localization enabled by VOEvents (\citealt{2023ATel16191....1R}). The coming years promise more interesting follow-up results from the ever-growing network of subscribers to the CHIME/FRB VOEvent Service. As of June 2024, the subscriber base consists of dozens of research institutions across four continents, including undergraduate students, graduate students, postdoctoral fellows, and faculty members, together numbering in the several hundreds. 

This paper and the associated open-source codebase are meant to provide a starting point for the community. It is hoped that collaboration with interested parties conducting multi-wavelength and/or multi-messenger follow-up will help to refine the FRB VOEvent format that \texttt{frb-voe} uses. It would be useful to have non-radio observatories producing VOEvents about their FRB follow-up. For instance, a radio observatory detects an FRB, publishes a VOEvent, and an X-ray observatory follows up and publishes its own VOEvent, with a citation of the original VOEvent. Finally, an optical observatory may respond to the X-ray follow-up rather than the initial alert from the radio observatory.

\section{Conclusion \label{sec:conclude}}

In summary, this paper presents a low-latency public alert service for detections of new and repeating FRBs from the CHIME/FRB experiment, as well as a telescope-agnostic FRB VOEvent alert service. This public alert service is the first of its kind and has decreased FRB follow-up latencies by orders of magnitude, and has led to several multi-wavelength follow-up campaigns. We strongly encourage observatories with an FRB search or follow-up program to consider implementing their own public alert system, either by using \texttt{frb-voe} out of the box or forking the repository and customizing it for their needs. Importantly, \texttt{frb-voe} offers convenient tools to interact with the TNS for obtaining official names of FRB discoveries in a standardized way, which is a path towards simplifying and unifying global FRB searches. 

\section{Acknowledgements}
Development of the CHIME/FRB VOEvent and \texttt{frb-voe} services was co-led by Thomas C. Abbott and Andrew V. Zwaniga, with A.V.Z. creating the CHIME/FRB VOEvent Service and first version of \texttt{frb-voe} and T.C.A continuing the development of the CHIME/FRB VOEvent Service and creating the current version of \texttt{frb-voe}. T.C.A is supported by the Centre de recherche en astrophysique du Québec, un regroupement stratégique du FRQNT. A.V.Z. is supported by an Ontario Graduate Scholarship. V.M.K. holds the Lorne Trottier Chair in Astrophysics \& Cosmology, a Distinguished James McGill Professorship, and receives support from an NSERC Discovery grant (RGPIN 228738-13), and from CIFAR. M.B. is a Mcwilliams fellow. A.M.C. is funded by an NSERC Doctoral Postgraduate Scholarship. K.W.M. holds the Adam J. Burgasser Chair in Astrophysics. A.P. is funded by the NSERC Canada Graduate Scholarships -- Doctoral program. Z.P. is supported by an NWO Veni fellowship (VI.Veni.222.295). K.S. is supported by the NSF Graduate Research Fellowship Program. S.P.T. is a CIFAR Azrieli Global Scholar in the Gravity and Extreme Universe Program. Figure \ref{fig:detection}: Host Observatory (``Parabolic Dish" by Vectors Point, from Noun Project, CCBY3.0),  frb-voe Server (``Server" by Mick Apps, from Noun Project, CCBY3.0), frb-voe operator (``Human" by nesibefeos, from Noun Project, CCBY3.0). 

\bibliography{sample631}{}
\bibliographystyle{aasjournal}

\end{document}